\documentclass[a4paper,11pt]{article}

\usepackage{jcappub} 

\usepackage[T1]{fontenc} 
\usepackage{amsmath}
\usepackage{graphicx} 

\usepackage{array}
\usepackage{xcolor}
\usepackage{cancel}
\usepackage[utf8x]{inputenc}
\usepackage[left]{lineno}
\usepackage[normalem]{ulem}
\usepackage{lineno}
\title{Fermionic versus Bosonic Dark Matter in Neutron Stars: A Bayesian Study with Multi-Density Constraints}

\author{Payaswinee Arvikar \textsuperscript{1,2}}
\author{Sakshi Gautam \textsuperscript{3}}
\author{Anagh Venneti \textsuperscript{2}}
\author{Sarmistha Banik \textsuperscript{2}}

\affiliation{\textsuperscript{1}Dharampeth M. P. Deo Memorial Science College, Nagpur 440033, India}
\affiliation{\textsuperscript{2}Department of Physics, BITS-Pilani Hyderabad Campus, Hyderabad, 500078, India}
\affiliation{\textsuperscript{3}Department of Physics, Panjab University, Chandigarh, 160014, India}

\emailAdd{p20230533@hyderabad.bits-pilani.ac.in}
\emailAdd{sakshigautam89@gmail.com}
\emailAdd{p20210060@hyderabad.bits-pilani.ac.in}
\emailAdd{sarmistha.banik@hyderabad.bits-pilani.ac.in}

\abstract{We perform a comparative Bayesian analysis of fermionic and bosonic dark matter admixed neutron stars (DMANS) by incorporating a comprehensive set of theoretical, experimental, and astrophysical constraints. The hadronic matter equation of state (EoS) is modeled using a relativistic mean-field approach, constrained by chiral effective field theory ($\chi$EFT) calculations at low densities, finite nuclei and heavy-ion collision data at intermediate densities, and neutron star (NS) observations at high densities. For the dark sector, we consider fermionic dark matter (FDM) interacting via a dark vector meson, and two bosonic dark matter models (BDM1 and BDM2) characterized by self-interacting scalar fields. Bayesian inference is employed to constrain the model parameters, including the dark matter mass, coupling strength, and dark matter fraction within NSs. Our analysis finds that all models yield consistent nuclear matter parameters, allowing a small dark matter fraction under 10\%. The presence of dark matter slightly softens the EoS, leading to a modest reduction in NS  mass, radius, and tidal deformability, though all models remain compatible with NICER and GW170817 observations. The log-evidence and likelihood analyses reveal no statistical preference among the FDM and BDM models, indicating that current astrophysical data cannot decisively distinguish between fermionic and bosonic dark matter scenarios. This study provides a unified statistical framework to constrain dark matter properties using NS observables.}
\keywords{Bayesian reasoning, dark matter theory, neutron star}

\begin{document}
\maketitle
\flushbottom

\section{Introduction}
Dark matter (DM), comprising about 27\% of the Universe's energy density, remains one of modern physics' greatest mysteries. While it plays a crucial role in cosmic structure and galactic dynamics, its particle nature is unknown since, lacking electromagnetic interactions, it reveals itself only through gravitational and indirect signatures.
Over the past several decades, a broad spectrum of dark matter candidates has been proposed. Among fermionic candidates, weakly interacting massive particles (WIMPs), sterile neutrinos, and gravitinos have received considerable attention \cite{WIMP_Bertone_2018,Dmreview_Bertone_2018}. In contrast, bosonic candidates encompass axions, axion-like particles (ALPs), and ultralight scalar fields \cite{Axion_Duffy_2009,Klangburam_2025}. Each class of candidates has distinct implications for structure formation, cosmic evolution, and astrophysical signatures. Consequently, constraining the parameter space of both fermionic and bosonic dark matter remains a crucial step toward unveiling its true nature.
\par
The extreme densities and strong gravitational fields in neutron stars (NSs) enhance DM capture, accumulation, and possible self-interactions, offering new avenues to constrain both fermionic and bosonic scenarios. The amount of dark matter present in a NS depends on the star’s evolutionary history as well as its location over its lifetime. Depending on the accumulated quantity, DM may concentrate in the center of the star to form a DM core \cite{Ivanytskyi_2020,Giangrandi_2024,AnkitKumar_2024}, or it may spread beyond the baryonic radius, giving rise to a DM halo \cite{Karkevandi2021}. The presence of DM can affect the space-time around NS, and thus observable properties such as mass, radius, tidal deformabilities and thermal evolution etc. get influenced.
\par
For instance, studies have investigated fermionic DM capture and its effects on NS heating and stability \cite{Kouvaris_2010,Lavallaz_2010}, while others have explored the possibility of bosonic DM forming condensates inside compact stars \cite{Karkevandi_2024,Husain_2021,Grippa_2024}, altering their EoS and leading to exotic stellar configurations.
Fermionic DM can provide additional pressure support, helping to stabilize the star against gravitational collapse. However, at very high densities and low temperatures within the core, bosonic DM may undergo Bose–Einstein condensation (BEC).
If the mass of dark matter admixed neutron stars (DMANS) exceeds a critical limit, the presence of a BEC can trigger gravitational collapse into a black hole. The stability of such BE condensates can be maintained by introducing a repulsive self-interaction, often modeled through a vector field coupled to the DM particle field.
\par
Numerous studies have explored the presence of fermionic and bosonic dark matter in the interiors of compact stars, with a few works focusing specifically on contrasting the two cases. For example, 
The properties of DMANSs were examined in \cite{Leung_2022}, where dark matter was modeled as either a simple free Fermi gas or a bosonic scalar field, and the tidal characteristics of the stars were analyzed.
Maselli {\it et al}. \cite{Maselli_2017} investigated rotating dark stars and showed that the I-Love-Q universal relations hold for both fermionic and bosonic dark stars. Ellis {\it et al.} \cite{Ellis_2018} studied the fermionic and bosonic DM models with the anticipation that once enough positive pressure is created at the centre of a star to prevent it from collapsing to a black hole, either kind of DM would show similar trends in the properties. Husain and Thomas \cite{Husain_2021} found that the tidal deformability constraint for a 1.4 solar mass DMANS permits up to about 5–10\% fermionic dark matter and up to 18\% bosonic dark matter in the stellar interior. In contrast, Karkevandi {\it et al.} \cite{Karkevandi2021} argued that the combined requirement of supporting two-solar-mass NS and satisfying LIGO/Virgo tidal deformability bounds restricts the bosonic dark matter fraction to below 5\%. However, a follow-up study \cite{Karkevandi_2024} demonstrated that for moderately heavier bosons, the permissible dark matter fraction may rise to about 20\%. In a related direction, Rutherford {\it et al.} \cite{Rutherford_2023} demonstrated that the forthcoming STROBE-X mission is expected to provide more stringent constraints on the bosonic dark matter parameter space than NICER, particularly once uncertainties in the baryonic EoS are reduced. 
Recently, the same collaboration used NS mass-radius measurements to investigate fermionic asymmetric dark matter \cite{Rutherford_2025} finding that meaningful constraints can be placed on the dark matter’s particle mass and the strength of its effective self-repulsion. Zhang et al. \cite{Zhang_2025} recently 
advocated a DM fraction of less than 2\% using simultaneous mass-radius measurements of PSR J0740+6620. Another study of Stubbs {\it et al.} \cite{Buras_Stubbs_2024} investigated the parameter space of dark matter using hybrid NS embedded in anisotropic bosonic dark matter halo supplemented with observational constraints from dynamics of galaxy clusters. 
Thus, it can be inferred that constraints on the dark matter parameter space remain under debate, as they depend strongly on the assumed dark matter particle type and on the astrophysical observations of compact stars used to limit them.\cite{DelPopolo:2019, Ivanytskyi_2020,Karkevandi2021,Shakeri_2022, Routaray_2023, Deliyergiyev:2023, Giangrandi_2024,Guha:2024, Ellis_2018,Zhang_2025, Liu:2025}. 
\par
The true particle nature of dark matter—fermionic or bosonic—remains an open question, with both possibilities actively pursued within the scientific community. In this study, we aim to address these competing hypotheses  through a comparative analysis of fermionic and bosonic DMANS. Here, we attempt for the first time a comparison between fermionic and bosonic DMANS, employing Bayesian optimization considering similar uncertainties in the RMF hadronic sector. By investigating how different dark matter candidates imprint on observable stellar properties, we assess which scenario aligns more closely with current astrophysical data. This approach does not presume the dominance of either particle type but rather provides a framework to critically evaluate their relative plausibility in light of observations.
\par
In our previous investigation \cite{Arvikar:2025}, we studied the properties of DMANS for fermionic DM and observed that current astrophysical observations could only constrain the DM fraction, which is primarily governed by EoS of hadronic matter and is weakly influenced by the method of incorporating astrophysical constraints. The hadronic EoS is constructed to satisfy constraints from finite nuclear properties, while also incorporating experimental results from heavy-ion collisions and astrophysical observations of NS. Notably Chiral Effective Field Theory ($\chi$EFT) calculations were not incorporated in that study. 
In the present study, we extend  our analysis to compare the properties of NSs for fermionic and bosonic DM.  For this purpose, we employ Bayesian inference on the hadronic and dark-matter EoS by incorporating constraints spanning  a broader density range: (a) low-density constraints from $\chi$EFT calculations \cite{Keller2023}, (b) intermediate-density constraints from finite nuclei properties and heavy-ion collisions data \cite{Tsang:2020,Venneti_2024} and (c) high-density constraints  as per astrophysical observations of NS \cite{Riley_2019,Riley_2021, Miller_2019,Miller_2021,Choudhury:2024xbk,Mauviard:2025,Abbott_2018}. 
This framework allows us to compare the parameter space of fermionic and bosonic dark matter candidates and to highlight how their distinct physical characteristics manifest under these multi-scale constraints.

\section{Equation of state for nuclear matter and dark matter}\label{EoS HM DM}
\subsection{Hadronic Matter Equation of State}

The Lagrangian density for the RMF model used in the present work is based upon different non-linear, self and inter-couplings among isoscalar-scalar $\sigma$, isoscalar-vector $\omega$,  and isovector-vector $\rho$ meson fields and nucleonic Dirac field $\psi$ and is given by \cite{Dutra_2014, Venneti_2024, SGAUTAM_2024},

\begin{eqnarray}\label{eq:lag}
    \nonumber
   \mathcal{L}&=& \!\bar{\psi}\left(i\gamma^\mu\partial_\mu - M\right)\psi +\!\!g_\sigma \sigma \bar{\psi} \psi -\!\!g_\omega \bar{\psi}\gamma^\mu\omega_\mu\psi -\frac{g_\rho}{2}\bar{\psi}\gamma^\mu\vec{\rho}_\mu +\frac{1}{2}\left(\partial^\mu\sigma\partial_\mu\sigma - m^2_\sigma \sigma^2\right) \\ \nonumber
   &&- \frac{\text{A}}{3} \sigma^3 - \frac{\text{B}}{4} \sigma^4 \! -\frac{1}{4} \Omega^{\mu\nu}\Omega_{\mu\nu} + \frac{1}{2}m^2_\omega\omega^\mu\omega_\mu +\frac{\text{C}}{4}\left(g_\omega^2 \omega_\mu \omega^\mu\right)^2 \\ 
   &&-\frac{1}{4}\vec{B}^{\mu\nu}\vec{B}_{\mu\nu} + \frac{1}{2}m^2_\rho \vec{\rho}_\mu \vec{\rho}^{ \mu} + \frac{1}{2} \Lambda_v g^2_\omega g^2_\rho \omega_\mu \omega^\mu \vec{\rho}_\mu \vec{\rho}^\mu. \\ \nonumber
\end{eqnarray}    
In the above equations, $\Omega_{\mu\nu} = \partial_\nu \omega_\mu - \partial_\mu \omega_\nu$ and $\vec{B}_{\mu\nu} = \partial_\nu \vec{\rho}_\mu - \partial_\mu \vec{\rho}_\nu - g_\rho \left(\vec{\rho}_\mu \times \vec{\rho}_\nu \right)$.
In the mean-field approximation, the mesonic fields are replaced by their expectation values, $\sigma \rightarrow \langle \sigma \rangle\equiv \sigma, \omega \rightarrow \langle\omega_\mu\rangle \equiv \omega_0 $  and $\bar{\rho}_\mu \rightarrow\langle\bar\rho_\mu\rangle\equiv\bar\rho_{0(3)}$ in the ground state. The field equations are solved to obtain the energy density and pressure values of the nuclear matter, which are given below \cite{Venneti_2024}.

\begin{eqnarray}{\label{eos_nm}}
   \mathcal{E}_{HM} = \frac{1}{\pi^2} \int_{0}^{k_p,k_n} dk k^2 \sqrt{k^2 + (M^*)^2}  + \frac{1}{2} m_\sigma^2 \sigma^2 - \frac{1}{2} m_\omega^2 \omega^2 -  \frac{1}{2} m_\rho^2 \rho^2 + \frac{\text{A}}{3} \sigma^3 + \frac{\text{B}}{4} \sigma^4 \nonumber \\ \nonumber
   - \frac{\text{C}}{4} g_\omega^4 \omega^4 + g_\omega \omega(\rho_p + \rho_n) - \frac{\Lambda_v}{2} (g_\rho g_\omega \rho \omega )^2 + \frac{g_\rho}{2} \rho (\rho_p - \rho_n), \\ \nonumber
   P_{HM} = \frac{1}{3 \pi^2} \int_{0}^{k_p,k_n} dk \frac{k^4} {\sqrt{k^2 + (M^*)^2}} - \frac{1}{2} m_\sigma^2 \sigma^2 + \frac{1}{2} m_\omega^2 \omega^2 +  \frac{1}{2} m_\rho^2 \rho^2 - \frac{\text{A}}{3} \sigma^3 - \frac{\text{B}}{4} \sigma^4 \\ 
   + \frac{\text{C}}{4} g_\omega^4 \omega^4 +  \frac{\Lambda_v}{2} (g_\rho g_\omega \rho \omega )^2. \\ \nonumber
\end{eqnarray}
In the above equations, $\rho_{p,n}$ and $k_{p,n}$ represent the density and the Fermi momentum of protons or neutrons and $M^{*}=M-g_{\sigma}\sigma $ represents the effective nucleonic mass.

\subsection{Fermionic Dark Matter Equation of State}

\noindent For fermionic dark matter, we follow our earlier approach \cite{Arvikar:2025}, where DM particles ($\chi_D$) are considered to be fermions with mass $M_D$. A `dark vector meson' ($V_D^\mu$) of mass $m_{vd}$ couples to DM particle via coupling strength $g_{vd}$. The Lagrangian density for the DM sector is,

\begin{equation}
    \mathcal{L}_\chi = \!\bar{\chi}_D [\gamma_\mu (i\partial^\mu - g_{vd} V_D^\mu ) - M_D ] \chi_D - \frac{1}{4} V_{\mu\nu,D} V_D^{\mu\nu} + \frac{1}{2} m^2_{vd} V_{\mu,D} V_D^\mu.
\end{equation}
Following this, the energy density and pressure for DM are given as \cite{Thakur_2024},

\begin{eqnarray}{\label{eos_dm1}}
    \mathcal{E}_D &=& \frac{1}{\pi^2} \int_{0}^{k} dk~ k^2 \sqrt{k^2 + M_D^2} + \frac{g_{vd}^2}{2m_{vd}^2} \rho_D^2, \nonumber \\  
    P_D &=& \frac{1}{3 \pi^2} \int_{0}^{k} dk \frac{k^4} {\sqrt{k^2 + M_D^2}} + \frac{g_{vd}^2}{2m_{vd}^2} \rho_D^2 ,
\end{eqnarray}
where $k$ is the Fermi momentum for DM. The ratio $C_{vd} = g_{vd}/m_{vd}$ is for vector interaction between DM particles and dark mesons. $\rho_D$ represents the density of DM fermions associated with the mean field value of dark vector mesons \cite{thakur2023exploring}. Hereafter, we refer to this model as FDM.

\subsection{Bosonic Dark Matter Equation of State}

\emph{Bosonic DM Model 1 (BDM1)}:\\
In this model, dark matter is described as a system of massive, self-interacting bosons, represented by a complex scalar field with an associated self-interaction potential. The linearized mean-field Lagrangian for such a DM with mass $M_D$ is given by \cite{Karkevandi2021,Karkevandi_2024}, 

\begin{equation}{\label{BDM1_L}}
    L = \frac{1}{2}\partial_\mu\phi^*\partial^\mu\phi - \frac{M_D^2}{2}\phi^*\phi + \frac{\lambda}{4}<\phi^*\phi>^2.
\end{equation}
The first two terms in Eq. (\ref{BDM1_L}) represent the standard kinetic and mass contributions of the scalar field, while the third term accounts for self-interactions, with $\lambda$ denoting the dimensionless coupling constant. The Klein–Gordon–Einstein equations for spherically symmetric scalar fields were first solved in Ref. \cite{Colpi_1986}. At sufficiently low temperatures, this system can undergo Bose–Einstein condensation. The pressure $P_D$ and energy density $\mathcal{E}_D$ for this bosonic dark matter model are derived from the Lagrangian (See the appendix of Ref. \cite{Karkevandi2021}), and is given by
\begin{equation}
    P_D = \frac{M_D^4}{9\lambda}\bigg(\sqrt{1+\frac{3\lambda\mathcal{E}_D}{M_D^4}}-1\bigg)^2.
\end{equation}

This EoS has been shown to produce boson stars with masses high enough to be consistent with typical neutron stars described by hadronic EoSs.
It is worth mentioning that the choice of priors of DM parameters, {\it viz}. mass and coupling constant is motivated by Ref. \cite{Karkevandi_2024}. The results for this model are labeled by BDM1. 

\emph{Bosonic DM Model 2 (BDM2)}:\\
Another approach to consider anisotropic bosonic dark matter is presented  in Ref. \cite{Buras_Stubbs_2024} where a hybrid NS is  considered within an anisotropic dark matter halo. The Lagrangian is similar to that of  the previous model.
The rotational motion of the condensate is negligibly small and following the methodology in Bohmer \& Harko \cite{Boehmer:2007}, the dynamics of the condensate is reduced to a polytrope with the following EoS,
\begin{equation}
   P_D = \frac{2\pi l_D}{M_D^3} \mathcal{E}_D^2,
\end{equation}
where $l_D$ is the inter particle scattering length, an important parameter describing the DM Bose-Einstein condensate. Here, the choice of priors for the DM parameters—namely, mass and scattering length—is guided by Ref. \cite{Buras_Stubbs_2024}.

\section{Methodology}

\subsection{Two-fluid formalism}

The bosonic and fermionic DM models considered in this analysis favor only gravitational interaction between DM and HM. Hence, the Lagrangians for both the fluids are independent. With the HM and DM EoSs in chemical equilibrium, the mass and radius of NS is calculated solving the Tolman-Oppenheimer-Volkoff (TOV) equations for the two-fluid system given by,

\begin{eqnarray}{\label{tov}}
    \frac{dP_{HM}}{dr} &=& -(P_{HM}+\mathcal{E}_{HM}) \frac{4 \pi r^3 (P_{HM}+P_{DM}) + m(r)}{r(r-2m(r))} ,\nonumber\\ 
    \frac{dP_{DM}}{dr} &=& -(P_{DM}+\mathcal{E}_{DM}) \frac{4 \pi r^3 (P_{HM}+P_{DM}) + m(r)}{r(r-2m(r))} ,\nonumber \\ 
    \frac{dm (r)}{dr} &=& 4 \pi (\mathcal{E}_{HM}(r) + \mathcal{E}_{DM}(r)) r^2 .
\end{eqnarray}
where $m(r)$ is the mass enclosed in radius $r$ and $P_{HM}(\mathcal{E}_{HM}$) and $P_{DM}(\mathcal{E}_{DM}$) are pressure (energy density) of hadronic and dark matter. We define the DM fraction in DMANS as the ratio of the DM mass to the total mass of the NS, which allows us to control the amount of dark matter present in the system. i.e,  $f_{DM}= \frac{M_{DM}}{M_{Total}}$ \cite{ Sagun_2022,Routaray_2023, thakur2023exploring,Thakur_2024}. 
In the above expression, $M_{DM}$ is the mass enclosed by a radius $R_{DM}$, and is given by, 
\begin{equation}
M_{DM}(R_{DM}) = 4 \pi \int_0^{R_{DM}} r^2 \mathcal{E}_{DM}(r) dr. 
\end{equation}
The total mass of the star is calculated by,
\begin{equation}
M_{Total}(R_T) = 4 \pi \int_0^{R_T} r^2 (\mathcal{E}_{HM}(r)+\mathcal{E}_{DM}(r)) dr.
\end{equation}
From the TOV calculations, we calculate the total radius $R_T$ as well as the hadronic radius $R_H$ where the nuclear matter pressure vanishes. The difference between $R_T$ and $R_H$ indicates the possible extension of dark matter beyond the visible stellar radius. During the final stages of inspiral in a binary NS system, the tidal gravitational field generated by each companion induces quadrupole deformations in NS. These tidal forces lead to distortions whose magnitude is characterized by a parameter known as tidal deformability, quantified as,
$\lambda = \frac{2}{3} k_2 R_T^5$ and the dimensionless tidal deformability is given by $\Lambda = \frac{2}{3} k_2 \mathcal{C}^{-5}$. Here compactness $\mathcal{C} = M/R_T$ and $k_2$ is the tidal love number, the definitions can be seen in Ref. \cite{Arvikar:2025}.

\subsection{Constraints on the Equation of State}

In this comprehensive analysis, we incorporate theoretical, experimental, and observational constraints spanning different density regimes while studying the dark matter admixture in neutron stars.\\ 

\emph{Constraints at low densities}:
The theoretical calculations of the chiral effective field theory ($\chi$EFT) put very strict constraints on the low-density RMF hadronic EoSs. We incorporate the $\chi$EFT-based constraints by using the pressure values of the $\beta$-equilibrated EoS at low densities (0.1, 0.15, and 0.2 fm$^{-3}$), as reported in Ref. \cite{Keller2023}, to constrain our Bayesian-generated hadronic matter EoS.\\

\emph{Constraints based on finite nuclei and HIC data}:
We place constraints on the symmetry energy $J(\rho)$, pressure of symmetric nuclear matter $P_{SNM}(\rho)$ and the symmetry pressure $P_{sym}(\rho)$ \cite{Tsang:2020,Venneti_2024}. These constraints are derived from various experiments on either finite-nuclei (FNC) or heavy-ion collisions (HIC). For example, the analyses of dipole polarizability of $^{208}$Pb, isospin diffusion data and single ratios of neutron-proton spectra provide us with empirical constraints on $J(\rho)$ at densities $<0.5\rho_0$. At densities $0.5-1.5 \rho_0$, analyses of nuclear masses, PREX-II, and pion spectra from HICs have placed constraints on $J(\rho)$ and $P_{sym}(\rho)$. The ISGMR places limits on the incompressibility coefficient ($K_0$). $P_{SNM}(\rho)$ and $P_{sym}(\rho)$ are constrained by HICs at densities $2\rho_0$ and $\sim 1.5\rho_0$. Please refer to Refs. \cite{,Tsang:2020,Venneti_2024} for details.\\

\emph{Constraints based on Astrophysical observations}:
Furthermore, simultaneous mass-radius measurements using X-ray hot spots on the NS surface from the NICER mission provide us with additional constraints to be placed on the NS properties. 
Here, all the current NICER data is incorporated. The mass-radius data for the observations of pulsars PSR J0030+0451 \cite{Riley_2019,Miller_2019}, PSR J0740+6620 \cite{Riley_2021,Miller_2021}, PSR J0437-4715 \cite{Choudhury:2024xbk} and PSR J0614-3329 \cite{Mauviard:2025} are imposed as the astrophysical constraints along with the GW170817 \cite{Abbott_2018} observation. This study presents the first simultaneous incorporation of NICER data from four pulsars into Bayesian inference, enabling the optimization of fermionic and bosonic dark matter parameters.

\begin{table*}[t]
    \centering
    \renewcommand{\arraystretch}{2.5}
    \begin{tabular}{|c|c|c|c|c|c|}
        \hline
        $\rho_0$ (fm$^{-3})$ & E$_0$ (MeV)& K$_0$ (MeV)& $M^*/M$ & J$_0$ (MeV)& L$_0$ (MeV)\\
        \hline
        0.14-0.17 & -16.5 - (-15.5) & 150 - 300 & 0.5 - 0.8 & 20 - 40 & 20 - 100 \\
        \hline
    \end{tabular}
    \caption{Uniform priors defined for all NM parameters.} 
    \label{tab:NMP_priors}
\end{table*}

\begin{table*}[t]
    \centering
    \renewcommand{\arraystretch}{2.7}
    \begin{tabular}{|c|c|c|c|}
        \hline
        Model & Interaction parameter & $M_D$ (MeV) & $f_D (\%)$ \\
        \hline
        FDM & $C_{vd}$ = 0.0005 - 0.040 (MeV)$^{-1}$& 500 - 10$^4$ & 0 - 25 \\
        \hline
        BDM1 & $\lambda$ = 0.3 - 20 & 50 - 10$^3$ & 0 - 25 \\
        \hline
        BDM2 & log~$l_D$ = log(1.1 fm) - log(10$^6$ fm) & 500 - 10$^4$ & 0 - 25 \\
        \hline
    \end{tabular}
    \caption{Uniform priors defined for all DM parameters.}
    \label{tab:DMP_priors}
\end{table*}

\subsection{Bayesian Inference}

We employ a Bayesian statistical inference framework for the estimation of the DM parameters which utilizes Bayes' theorem \cite{stuart1994kendalls}, 
\begin{equation}\label{bayes}
    P(\mathbf{\Theta_M}|D) = \frac{\mathcal{L}(D|\mathbf{\Theta_M})P(\mathbf{\Theta_M})}{\mathcal{Z}},
\end{equation}
where P$(\mathbf{\Theta_M}|D)$ is the posterior probability for the model $\mathbf{M}$ with parameters $\mathbf{\Theta}$ ($\rho_0, E_0, K_0, \frac{M^*}{M}, \\ J_0, L_0,C_{vd}/\lambda/l_D, M_D, f_{DM}$) given the data set $D$, $\mathcal{L}(D|\mathbf{\Theta_M})$ is the likelihood function or conditional probability for a given theoretical model $\mathbf{M}$ to correctly predict the data $D$, and $P(\mathbf{\Theta_M})$ is the prior probability of the model $\mathbf{M}$ before confronting the data. The evidence $\mathcal{Z}$ is the normalization constant.

We use the Gaussian likelihood for the set of finite nuclei properties $P_{SNM}$, $K_0$, $J(\rho)$ and $P_{sym}(\rho)$ listed in table S3 of  Ref.\cite{Venneti_2024} and as mentioned in Eqn. \ref{gaus_L}. This likelihood function is also employed for the $\chi$EFT constraints.

\begin{equation}\label{gaus_L}
    \mathcal{L}_{\sigma}(D|\mathbf{\Theta}) = \frac{1}{\sqrt{2 \pi \sigma^2}} \exp \biggl ( -\frac{(D(\mathbf{\Theta})-D_{Obs})^2}{2 \sigma^2}\biggr).
\end{equation}
We use Kernel Density Estimator (KDE) in our likelihood to use the entire NICER and GW posterior datasets available. Here, the NICER data for the observations of PSR J0030+0451 \cite{Riley_2019,Miller_2019}, PSR J0740+6620 \cite{Riley_2021,Miller_2021}, PSR J0437-4715 \cite{Choudhury:2024xbk} and PSR J0614-3329 \cite{Mauviard:2025} are imposed to calculate $\mathcal{L}_{Obs}(D|\mathbf{\Theta})$.  The total likelihood $\mathcal{L}_{Obs}(D|\mathbf{\Theta})$ becomes
\begin{equation}
\mathcal{L}_{Obs}(D|\mathbf{\Theta}) = \mathcal{L}_{NICER}^{0740} \times \mathcal{L}_{NICER}^{0030} \times \mathcal{L}_{NICER}^{0437} \times \mathcal{L}_{NICER}^{0614}\nonumber,
\end{equation}
where
\begin{equation}
    \mathcal{L}_{NICER}(D|\mathbf{\Theta}) = \int^{M_{max}}_{M_0} dm ~ P(m|\mathbf{\Theta}) \times P(D|m,R(m,\mathbf{\Theta})).
\end{equation}
Using this probability, we calculate the likelihood for all the pulsar observations. Also, for GW observation of the binary system, $m_1, m_2$ are the masses and $\Lambda_1, \Lambda_2$ are the tidal deformabilities of the binary components.
\begin{multline}
    \mathcal{L}_{GW}(D|\mathbf{\Theta}) = \int^{M_u}_{m_l}dm_1 \int^{m_1}_{M_l} dm_2 ~ P(m_1 , m_2|\mathbf{\Theta}) \\ 
    \times P (d_{GW} |m_1 , m_2 , \Lambda_1 (m_1 ,\mathbf{\Theta}), \Lambda_2 (m_2 , \mathbf{\Theta})).
\end{multline}
This gives the likelihood $L_{GW}$, where $P(m|\Theta)$ is written as,

\begin{equation}
    P(m|\mathbf{\Theta}) = \begin{cases}
        \frac{1}{M_{max}-M_0}  & \text{if}\;  M_0 \leq m \leq M_{max} \\ \nonumber
        0 & \text{else}
    \end{cases}
\end{equation}
For these calculations, $M_0$ is always taken to be 1M$_{\odot}$ and $M_{max}$ is the maximum mass of the NS for given set of parameters.

\section{Results and Discussions}

In this work, we model dark matter in DMANSs using both fermionic and bosonic DM EoS, that is subjected to a broad set of constraints across different density ranges. As mentioned earlier, unlike our earlier study, the present analysis incorporates the constraints from chiral effective field theory at low densities, which are known to play a crucial role in constraining beta-equilibrated EoS \cite{Keller2023,TMalik:2023}. We further impose constraints on nuclear matter parameters (NMPs), including the symmetry energy, pressure of symmetric nuclear matter, and symmetry pressure, informed by experimental data from nuclear reactions and related studies, as outlined in the section III B, along with the astrophysical constraints.

\begin{figure*}
    \centering
    \includegraphics[width=\linewidth]{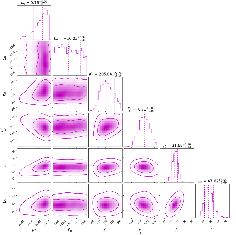}
    \caption{Posterior distributions of nuclear matter parameters (NMPs) for No DM case. The $1 \sigma$ CI is displayed as vertical dashed lines and median as dotted lines in the marginalized posterior distribution for the NMPs $\rho_0, E_0, K_0, \frac{M^*}{M},J_0,L_0$. Off-diagonal panels show the contours enclosing 1$\sigma$ and 2$\sigma$ CIs. All are in units of MeV, except for the saturation density $\rho_0$(fm$^{-3}$), and $M^*/M$ (dimensionless). }
    \label{fig:corner_NoDM}
\end{figure*}

We first perform the analysis without taking dark matter into consideration (labeled as `No DM'), where six NMPs are varied (as per priors listed in Table \ref{tab:NMP_priors}) and the median values of the NMPs for No DM case are listed in Table \ref{tab:posteriors}. The marginalized corner plots are shown in Fig.\ref{fig:corner_NoDM}. The diagonal panels display the marginalized posterior distributions for six NMPs, where vertical dashed lines mark the 1$\sigma$ confidence intervals (CIs). The off-diagonal panels illustrate the joint probability distributions, with contours enclosing the 1$\sigma$ and 2$\sigma$ CIs. We observe that all NMPs, with the exception of the saturation energy $E_0$, are well constrained. The compressibility coefficient $K_0$ is limited to around 200 MeV, depicting the softer nature of the EoS of symmetric nuclear matter.

Next, we add DM to nuclear matter and the resulting corner plots for the posteriors of the NMPs and DM EoS parameters are shown in Figs. \ref{fig:corner_FDM}-\ref{fig:corner_BDM2}. Specifically, Figs. \ref{fig:corner_FDM}, \ref{fig:corner_BDM1} and \ref{fig:corner_BDM2} present the results for fermionic (FDM) and bosonic (BDM1 and BDM2) DM EoSs respectively. Similar trend is observed in the constraining of NMPs across all the DM scenarios. For fermionic DM, the three DM parameters—coupling strength $C_{vd}$, DM particle mass $M_D$, and DM fraction $f_{DM}$—are only weakly constrained. The allowed DM fraction in DMANS, for this  model, is about 6.4\%. A similar trend is found for bosonic dark matter in the BDM1 model, where the interaction strength $\lambda$, $M_D$, and $f_{DM}$ remain poorly constrained. The DM fraction for bosonic DM is about 6.7\%. However, the FDM mass is limited to approximately 2700 MeV, while in the BDM1 model, with its shorter prior range, the DM particle mass is constrained to a lighter value of around 692 MeV. However, a higher DM fraction of about 8.9\% is obtained, with the DM boson mass constrained to roughly 1632 MeV. The posteriors of all NMPS along with DM parameters are listed in Table III. From the table, we infer the following:

\begin{figure*}
    \centering
    \includegraphics[width=\linewidth]{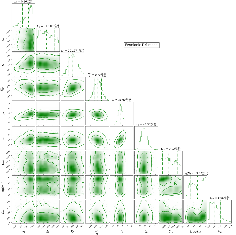}
    \caption{Posterior distributions of NMPs and DM parameters, $\rho_0, E_0, K_0, \frac{M^*}{M},J_0,L_0,C_{vd}, M_D, f_{DM}$, for fermionic DM model (FDM). Units of DM parameters are $C_{vd}$ (MeV)$^{-1}$ and $M_D$ (MeV). Contours and vertical lines represent CIs similar to those in Fig. \ref{fig:corner_NoDM}.}
    \label{fig:corner_FDM}
\end{figure*}

\begin{figure*}
    \centering
    \includegraphics[width=\linewidth]{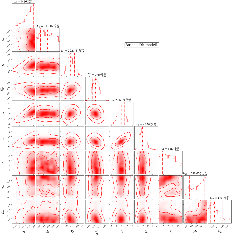}
    \caption{Posterior distributions of NMPs and DM parameters, $\rho_0, E_0, K_0, \frac{M^*}{M},J_0,L_0,\lambda, M_D, f_{DM}$, for first bosonic DM model (BDM1). Unit of $M_D$ is MeV and $\lambda$ is dimensionless. Contours and vertical lines represent CIs similar to those in Fig. \ref{fig:corner_NoDM}.}
    \label{fig:corner_BDM1} 
\end{figure*}

\begin{figure*}
    \centering
    \includegraphics[width=\linewidth]{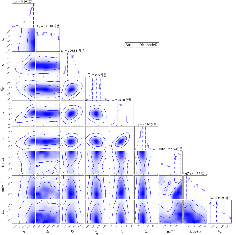}
    \caption{Posterior distributions of NMPs and DM parameters, $\rho_0, E_0, K_0, \frac{M^*}{M},J_0,L_0,l_D, M_D, f_{DM}$, for second bosonic DM model (BDM2). Units of DM parameters are $l_D$ (fm) and $M_D$ (MeV). Contours and vertical lines represent CIs similar to those in Fig. \ref{fig:corner_NoDM}.}
    \label{fig:corner_BDM2}
\end{figure*}

\begin{enumerate}
    \item The values of NMPs remain consistent across all DM models as well as for No DM case, primarily because they are constrained by the same $\chi$EFT calculations and FNC+HIC inputs.
    \item The fermionic DM model yields comparatively lower coupling values than the bosonic models. This distinction arises from the presence of Fermi pressure in the FDM case, which helps counterbalance gravitational collapse. In contrast, the absence of such pressure in bosonic DM necessitates stronger repulsive interactions to stabilize the system. 
    \item Comparable DM fractions are obtained in the FDM and BDM models, despite noticeable differences in the posterior distributions of coupling strengths and DM masses. This is explained by the interplay between mass and interaction strength that governs the DM fraction in DMANS. Specifically, the BDM model favors stronger couplings and lighter DM particles, whereas the FDM model accommodates weaker couplings with heavier DM candidates, while both satisfy the astrophysical constraints.
    \item The BDM2 model permits a higher dark matter fraction of 8.9\%, compared to about 6.7\% in the BDM1 case. The broader range of interaction leads to higher values of the coupling parameter, which in turn favors heavier dark matter particles in order to satisfy astrophysical constraints. Consequently, the resulting dark matter fraction in DMANS reaches about 8.9\%.
    \item It is worth noting that in our earlier study with the same FDM framework, we reported a DM fraction of upto 6\% in DMANS. This differs from the present analysis where DM fraction is higher (i.e. upto 12\%, considering 1$\sigma$ spread) and may be likely due to the inclusion of additional constraints from $\chi$EFT-based EoS and two recent pulsar observations in the Bayesian analysis.
\end{enumerate}
 
\begin{table*}[t]
    \centering
    \renewcommand{\arraystretch}{2.7}
    \begin{tabular}{|c|c|c|c|c|}
        \hline
        Parameters & No DM & FDM & BDM1 & BDM2 \\ 
        \hline
        $\rho_0$ (fm$^{-3})$ & $0.16^{+0.0045}_{-0.0080}$ & $0.16^{+0.0046}_{-0.0083}$ & $0.16^{+0.0045}_{-0.0077}$ & $0.16^{+0.0026}_{-0.0027}$ \\ 
        \hline
        $E_0$ (MeV)& $-16.03^{+0.33}_{-0.31}$ & $-16.05^{+0.33}_{-0.29}$ & $-16.01^{+0.31}_{-0.30}$ & $-16.03^{+0.33}_{-0.30}$\\ 
        \hline
        $K_0$ (MeV)& $205.04^{+22.25}_{-18.25}$ & $201.27^{+21.40}_{-17.52}$ & $202.45^{+22.06}_{-17.86}$ & $206.38^{+21.13}_{-18.76}$ \\ 
        \hline
        $M^*/M$ & $0.70^{+0.029}_{-0.028}$ & $0.69^{+0.027}_{+0.027}$ & $0.69^{+0.026}_{-0.025}$ & $0.70^{+0.028}_{-0.028}$ \\ 
        \hline
        $J_0$ (MeV)& $31.68^{+1.16}_{-1.33}$ & $31.60^{+1.21}_{-1.32}$ & $31.74^{+1.30}_{-1.33}$ & $31.85^{+1.19}_{-1.34}$ \\ 
        \hline
        $L_0$ (MeV)& $47.61^{+6.53}_{-5.68}$ & $48.29^{+6.32}_{-5.53}$ & $48.98^{+6.90}_{-5.70}$ & $48.67^{+5.99}_{-5.77}$ \\ 
        \hline
        Interaction parameter   & -- & $C_{vd}= 0.016^{+0.014}_{-0.010}$ & $\lambda = 8.8^{+6.60}_{-5.66}$ & $log ~l_D = 4.032^{+1.38}_{-2.72} $ \\ 
        \hline
        log $M_D$ (MeV) & -- & $3.4314^{+0.41}_{-0.48}$ & $2.8401^{+0.41}_{-0.28}$ & $3.2126^{+0.44}_{-0.34}$ \\ 
        \hline
        $M_D$(MeV) & -- & 2700.2 & 691.99 & 1631.5 \\ 
        \hline
        $f_{DM}$ (\%) & -- & $6.44^{+6.41}_{-4.38}$ & $6.72^{+8.86}_{-4.34}$ & $8.9^{+8.7}_{-6.1}$ \\ 
        \hline
    \end{tabular}
    \caption{The posterior medians and 68\% CI of NM and DM parameters for No DM case and for the fermionic and bosonic models under consideration. }
    \label{tab:posteriors}
\end{table*}

\begin{figure*}
    \centering
    \includegraphics[width=\linewidth]{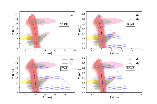}
    \caption{M-R distribution plots for pure NS [upper left panel] and DMANS with the three DM models [other three panels]. Red lines show all the M-R$_H$ curves obtained from the posteriors samples of Bayesian analysis. Blue hatched region is the 2$\sigma$ spread of the M-R distribution from the posteriors for total radius $R_T$ (hadronic + dark). Brown and blue dashed lines represent the corresponding medians for $R_H$ and $R_T$. NICER X-ray observations are shown as Pink and Green contours for PSR J0740+6620 )high mass) \cite{Miller_2021,Riley_2021} and PSR J0030+045 \cite{Miller_2019,Riley_2019}, respectively. Orange and Yellow contours correspond to the NICER data of pulsars PSR J0437-4715 \cite{Choudhury:2024xbk} and PSR J0614-3329 \cite{Mauviard:2025} respectively. The grey shaded M-R regions show 90\%(light) and 50\%(dark) CI for the LIGO/Virgo constraints derived from the binary components of GW170817 event.}
    \label{fig:tov_distri}
\end{figure*}

We further examine the properties of DMANS by solving the two-fluid TOV equations with Fig. \ref{fig:tov_distri} presenting the mass–radius (M–R) distributions derived from posterior samples obtained via Bayesian analysis.
The upper left panel shows the mass-radius curves for pure NS (No DM case) while the remaining three panels correspond to DMANS. 
NICER X-ray observations are shown as pink and green contours for PSR J0030+045 \cite{Miller_2019,Riley_2019} and PSR J0740+6620 (high mass) \cite{Riley_2021,Miller_2021}, respectively. Orange and yellow contours correspond to the NICER data of pulsars PSR J0437-4715 \cite{Choudhury:2024xbk} and PSR J0614-3329 \cite{Mauviard:2025}, respectively. The grey shaded M-R regions show 90\% (light) and 50\% (dark) CI for the LIGO/Virgo constraints derived from the binary components of the GW170817 event \cite{Abbott_2018}. The obtained M-R curves are shown by red lines and the medians are displayed by brown dashed lines.    
Note that this represents the median distribution corresponding to the hadronic (or nuclear) radius $R_H$.
The hatched regions represent the total radius $R_T$ (hadronic + dark). The median M–$R_T$ curves are  consistent with astrophysical constraints, overlapping with NICER X-ray data for pulsars as well as the GW170817 limits, which demonstrates that the inclusion of DM remains compatible with observations.
It is evident that a DM halo forms in both fermionic and bosonic cases, as the DM distribution extends beyond the visible stellar radius ($R_H$ is less than total $R_T$). Nevertheless, the median values of the visible and total radii remain close, indicating that DM is largely confined to the core, with only a small likelihood of forming a halo. 
A quantitative estimate shows that the probabilities of halo formation (defined as cases where $R_T > 1.1 \times R_H$) are approximately 9.6\%, 4.6\%, and 7.6\% for the FDM, BDM1, and BDM2 models, respectively. This indicates that halos are generally rare, occurring most frequently in FDM and least frequently in BDM1.
Our results further suggest that, for the BDM1 case, stars with masses exceeding 2 M$_\odot$ tend to contain DM cores instead of forming extended DM halos.

\begin{figure*}
    \centering
    \includegraphics[width=\linewidth]{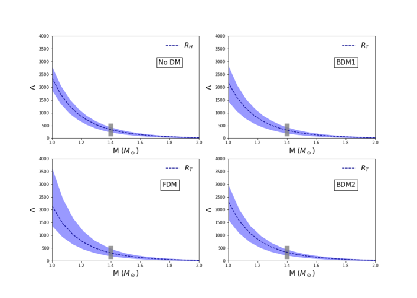}
    \caption{M-$\Lambda$ distribution plots for pure NS [upper left panel] and DMANS for three DM models [other three panels]. Blue dashed lines present the median distribution whereas the shaded region corresponds to 2$\sigma$ CI. The canonical tidal deformability $\Lambda_{1.4}$ for the observations of GW170817 event is displayed by grey band. }
    \label{fig:tov_distri2}
\end{figure*}

The dimensionless tidal deformability $\Lambda$ as a function of NS mass is presented in Fig. \ref{fig:tov_distri2}, for No DM (upper left panel) and all the DM models. The blue band shows the 2$\sigma$ CI along with the median represented by the dashed lines. These regions are consistent with the 1$\sigma$ band of the $\Lambda_{1.4}$ value from the GW170817 observational data, shown by the grey band. The median values of the NS properties, viz., maximum mass $M_{max}$, visible radius $R_{H1.4}$ and tidal deformability $\Lambda_{1.4}$ obtained for different DM models are listed in Table \ref{tab:NSprop}.
From the table, we infer that the maximum mass of DMANS is lower than that of pure NS, as expected due to the softening of the EOS in the presence of DM. A slight reduction is also observed in both $R_{1.4}$ and $\Lambda_{1.4}$ values of DMANS. 

Furthermore, we compute the log evidence values for the four models to quantitatively assess their relative statistical support. The obtained values are as follows: FDM: –58.19, BDM1: –58.25, and BDM2: –57.97. The small Bayes factors between any pair of these models indicate that none of them is statistically preferred over the others. Hence, all DM models are statistically indistinguishable based on their log evidence values. It is important to note that different observational and theoretical constraints have been incorporated into the Bayesian analysis. To further assess model performance, we evaluate the log-likelihood values for each experimental data set individually. The corresponding results for the FDM and BDM models are summarized in Table \ref{tab:loglike}. Overall, all DM models reproduce the $\chi$EFT predictions and NS observational constraints reasonably well. However, their agreement with the FNC+HIC data and the GW170817 event is comparatively weaker, indicating that these datasets impose more stringent constraints on the underlying DM interactions.

\begin{table*}
    \centering
    \renewcommand{\arraystretch}{2.3}
    \begin{tabular}{|c|c|c|c|}
        \hline
         Model & $M_{max} (M_\odot)$ & $R_{H1.4}$ (km) & $\Lambda_{1.4}$ \\
         \hline
         No DM & 2.20 & 12.23 & 337.40\\
         FDM & 2.15 & 12.06 & 304.60\\
         BDM1 & 2.16 & 12.12 & 312.94\\
         BDM2 & 2.18 & 12.18 & 329.29\\
         \hline
        \end{tabular}
    \caption{The posterior median values of NS properties viz. maximum mass, radius (visible) and tidal deformability of $1.4M_\odot$ NS obtained from analyses for No DM case and for different DM models.}
    \label{tab:NSprop}
\end{table*}

We further examine the correlations between the log-likelihood values ($\mathcal{L}^{i}_{NICER/GW}$) for individual pulsar observations shown in Fig. 7 across four models: no dark matter (No DM), fermionic DM (FDM), and two bosonic DM models (BDM1 and BDM2). Each $\mathcal{L}^{i}_{NICER/GW}$ indicates how closely the model’s posterior distribution for observation $i$ aligns with the observed median values. The correlations between these log likelihoods reveal how improving the model fit for one observation influences the fits for others.

In the No DM model, we find a strong anti-correlation between the log-likelihoods of PSR J0030+0451 and PSR J0614–3329 ($\sim$–0.66), with a similar trend for PSR J0437–4715. This means that, under nucleonic matter constraints, improving the fit to J0030 worsens the fit to J0614 (and J0437). Including dark matter substantially alters this pattern. For DMANS, the log-likelihood of J0030 becomes strongly positively correlated with that of J0614, increasing to 0.78 (0.91) for BDM1 (BDM2) and 0.93 for FDM, and shows a similar trend to J0437. Other inter-pulsar correlations also change, indicating that the DM components alter the overall consistency between the model predictions and observational data.

\begin{table}
    \centering
    \renewcommand{\arraystretch}{1.5}
    \begin{tabular}{|c|c|c|c|c|c|c|c|}
        \hline
        Model & $\chi$\small{EFT} & \small{FNC + HIC} & \small{PSR}  & \small{PSR}  & \small{PSR}  & \small{PSR} & \small{GW170817}\\
        & &  & \small{J0030+0451} & \small{J0740+6620} & \small{J0437-4715} & \small{J0614-3329} & \\
        \hline
        No DM & -1.46 & -29.98 & -1.81 & -2.57 & -0.90 & -1.52 & -14.11 \\
        FDM & -1.49 & -30.09 & -1.88 & -2.24 & -0.84 & -1.42 & -14.05 \\
        BDM1 & -1.50 & -30.05 & -1.90 & -2.25 & -0.80 & -1.35 & -14.02 \\
        BDM2 & -1.51 & -29.98 & -1.83 & -2.14 & -0.89 & -1.51 & -14.06 \\
        \hline
    \end{tabular}
    \caption{Log-likelihood values for individual data set for different DM models.}
    \label{tab:loglike}
\end{table}

\begin{figure*}
    \centering
    \includegraphics[width=0.5\linewidth]{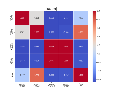}\includegraphics[width=0.5\linewidth]{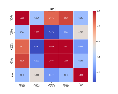}
    \includegraphics[width=0.5\linewidth]{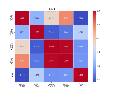}\includegraphics[width=0.5\linewidth]{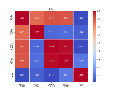}
    \caption{Correlation matrix for the Log-likelihood values with individual pulsars and GW constraints, for the No DM case and three DM models considered in the study.}
\end{figure*}

\section{Conclusion}

A comparative Bayesian study is conducted on fermionic and bosonic dark matter admixed neutron stars (DMANS) using a wide range of theoretical, experimental, and astrophysical constraints. The hadronic matter is modeled within the relativistic mean-field framework constrained by $\chi$EFT, nuclear, and astrophysical data. For the dark sector, fermionic dark matter with vector interactions and two self-interacting bosonic models are examined. We follow the two-fluid approach for the analyses. Bayesian inference constrains dark matter parameters, revealing that all models support consistent nuclear matter properties and allow a dark matter fraction below 10\%, with fermionic models favoring heavier and weakly interacting candidates and bosonic models favoring lighter and more strongly coupled particles. The inclusion of dark matter slightly softens the EoS, reducing neutron star mass, radius, and tidal deformability, yet remaining consistent with NICER and GW170817 observations. No statistical preference emerges among the fermionic and bosonic models, suggesting current data cannot distinguish between these dark matter scenarios. We may infer that future high-precision mass–radius and tidal deformability measurements will be crucial in resolving this degeneracy and refining constraints on dark matter properties inside compact stars.

\section*{Acknowledgment}
PA acknowledges the computing facilities at BITS Pilani-Hyderabad Campus, used for the analyses. The authors gratefully acknowledge fruitful discussions with Tuhin Malik. AV acknowledges the CSIR-HRDG for support through the CSIR-JRF 09/1026(16303)/2023-EMR-I.

\bibliographystyle{JHEP} 
\bibliography{DM_NS}
\end{document}